\documentclass[twocolumn]{revtex4}
\usepackage{graphicx}

\begin{document}

\title{Decoherence of multi-dimensional
entangled coherent states}
\author{S. J. van Enk}
\affiliation{Bell Labs, Lucent Technologies\\
600-700 Mountain Ave, Murray Hill, NJ 07974}
\date{\today}

\begin{abstract}
For entangled states of light 
both the amount of entanglement and
the sensitivity to noise generally 
increase with the number of photons in the state.
The entanglement-sensitivity
tradeoff is investigated for a particular
set of states, multi-dimensional entangled coherent states.
Those states possess an arbitrarily large amount of 
entanglement $E$
provided the number of photons is at least of order $2^{2E}$.
We calculate how fast that entanglement decays
due to photon absorption losses and how much entanglement is left. 
We find that for very small losses the amount of entanglement lost
is equal to $2/\log(2)\approx 2.89$ ebits per absorbed photon,
irrespective of the amount of pure-state entanglement $E$
one started with.
In contrast, for larger
losses it tends to be the remaining
amount of entanglement that is independent
of $E$.
This may provide a useful strategy for creating states 
with a fixed amount of entanglement.
\end{abstract}

\maketitle
\section{Introduction}
Entanglement lies at the heart of
many quantum information processing protocols \cite{nielsen}.
In order to generate entangled states 
one typically envisages a process that in the ideal case would
lead to the desired state with the desired amount of entanglement. 
Imperfections such as noise and losses, however, degrade the
quality of the state and diminish the amount of entanglement created.
As a consequence, the task for which the entangled state is created
(say, teleportation \cite{telep}) will be accomplished only partially.
An alternative strategy for generating a certain amount of entanglement 
would be the following: use a process that in the ideal case would 
generate (much) more entanglement than needed,
but that produces the desired amount in the realistic imperfect case.

If it were increasingly difficult to generate larger amounts of entanglement,
that strategy would not be very useful. Fortunately, it turns out 
that
for light beams
a standard optical nonlinearity \footnote{The Kerr nonlinearity 
arises, e.g., in nonlinear fibers: it 
has been known for a few decades that such a
nonlinearity can be used to
generate nonclassical states such as Schr\"{o}dinger cat states
\cite{yurke} and entangled coherent states \cite{sanders}.}
allows one to generate {\em more}
entanglement 
when the interaction time (or length) $\tau$ is {\em shorter}:
one generates $\log_2 M$ ebits in a time $\tau=1/M$ \cite{mdecs}.
This effect exploits some of the peculiar properties
of entanglement in infinite-dimensional Hilbert spaces \cite{infinite}.
We may use that effect
in the following way: take a very short time (or length) $\tau=1/M$ 
to generate a bipartite $M\times M$ dimensional
pure state of two modes with $\log_2 M$ ebits of entanglement, with $M$ large.
Since the interaction time is assumed short we may
neglect decoherence during 
that part of the procedure and consequently the state at that point
is still pure. 
By the time we will actually use the state
for some quantum-information processing protocol, there will have been
decoherence, turning the pure state
into a mixed state $\rho$. The most relevant type
of decoherence for light beams is that due to photon absorption losses,
and that is the only type of decoherence we will consider here.
We are interested in the 
amount of bipartite entanglement between the two modes \cite{modes}
in the state $\rho$. 

Since $\rho$ will typically still be a 
very high-dimensional state, we use a measure of entanglement that 
can actually
be computed efficiently even in that case: 
the logarithmic negativity $E_N(\rho)$ \cite{vidal}.
That quantity can be linked to two
operational measures of entanglement, demonstrating
the possible use of the mixed-state entanglement in $\rho$. On the one hand
$E_N(\rho)$ determines a bound (through the negativity in fact)
on the fidelity that can be achieved
in teleportation using a single copy of $\rho$, on the other it gives a bound
on the amount of distillable (pure-state) \cite{distill} entanglement
present in many copies of $\rho$ (see Ref.~\cite{vidal} for details).

An alternative 
method to create certain desired entangled states of light
was discussed in \cite{small}: that method
is very similar in spirit to that analyzed here 
but still relies on 
producing a pure state.
The first part of the procedure is the same as above
and consists of generating
an $M\times M$-dimensional pure entangled state (with $M$ large)
of two modes.
But subsequently one performs
appropriate local measurements on both modes 
to (approximately and probabilistically)
project 
the state down to 
a $2\times 2$-dimensional state of the form
\begin{equation}\label{ecs}
|\alpha\rangle\otimes|\alpha\rangle 
+\exp(i\phi) |-\alpha\rangle\otimes|-\alpha\rangle,
\end{equation}
where $|\alpha\rangle$ is a coherent state, and where the amplitude
$\alpha$ is not too small,
such that $|\alpha\rangle$ and 
$|-\alpha\rangle$ are to a very good approximation
orthogonal. In that case, the states (\ref{ecs})
possess one ebit of entanglement for any value of the phase $\phi$ and such
states can then be used for teleportation \cite{hirota}, 
quantum computing \cite{qcomp}, 
quantum communication \cite{qcom},
and testing nonlocality \cite{Bell}. 

Generally speaking,
the sensitivity to noise increases with the number of photons
in the state. This was noted, in particular,
in experiments
on generating Schr\"{o}dinger cat states of the form
$|\alpha\rangle+\exp(i\phi)|-\alpha\rangle$ \cite{haroche,myatt}.
It is also true for the related
entangled coherent states \cite{hirota,lixu},
and this issue was subsequently
investigated in more detail and generality 
for light beams in \cite{rob}.
For an even more general discussion of decoherence
of macroscopic and microscopic objects, see for example \cite{strunz}.
Thus, while increasing $M$ leads to more entanglement
it also leads to the state being more sensitive to decoherence, 
as the number of photons needed to create that much entanglement
increases as $M^2$. That is,
one expects the state to lose entanglement at a rate that increases with $M$.
Thus, one may expect an optimum
number of photons for fixed interaction time and fixed amount of decoherence
to exist that maximizes 
the amount of entanglement left 
after decoherence.
Confirming that expectation is another aim
of this paper.
\section{Analytical results}
\subsection{Multi-dimensional entangled coherent states}
We first review some of the results of 
\cite{mdecs} that we need here. 
Suppose one starts out with
a coherent state $|\sqrt{2}\alpha\rangle$ and one lets
the state propagate through a Kerr nonlinear medium, 
described by a Hamiltonian of the form $H=a^{\dagger 2}a^2$
for a
 time $\tau$ equal to 
$1/M$, with $M$ an integer.
Subsequently one splits the resulting state on a 50/50
beamsplitter with the vacuum.
The state of the two output modes so generated is of the form
\begin{equation}\label{M}
|\Phi\rangle=\sum_{q=1}^{M}f_q|\Phi_q\rangle
\otimes|\Phi_q\rangle,
\end{equation}
where the states $|\Phi_q\rangle$ are coherent
states with amplitude $\alpha$ and phases $-2\pi q/M$,
\begin{equation}\label{Phi}
|\Phi_q\rangle=|\alpha\exp(-2\pi iq/M)\rangle.
\end{equation}
The states (\ref{M}) generalize
the states (\ref{ecs}) and may be called
multi-dimensional entangled coherent states.
For ease of notation,
here and in the following we suppress the dependence
of the states $|\Phi_q\rangle$ and similar states defined below
on the parameters $\alpha$ and $M$.
Up to irrelevant
overall phase factors the coefficients $f_q$ are given by
\begin{eqnarray}\label{MM}
f_q&=&\frac{1}{\sqrt{M}}\exp\big(\frac{\pi iq(q+1)}{M}  \big),
\,\,\,M\,{\rm odd}\\
f_q&=&\frac{1}{\sqrt{M}}\exp\big(\frac{\pi iq^2}{M} \big)
,\,\,\,M\,{\rm even}.
\end{eqnarray}
The states (\ref{M}) potentially possess
$\log_2 M$ ebits of entanglement,
namely when $\alpha$ is sufficiently large to make the states
$|\alpha\exp(-2\pi iq/M)\rangle$
and
$|\alpha\exp(-2\pi iq'/M)\rangle$ nearly orthogonal for $q\neq q'$.
This requires that $|\alpha|^2$ be at least
of order $M^2$. More precisely,
if we choose a small number $\delta\ll 1$
the requirement
\begin{equation}
|\langle \Phi_q|\Phi_{q+1}\rangle|^2=\delta,
\end{equation}
gives
\begin{equation}\label{d}
|\alpha|^2\approx \frac{\log(1/\delta)}{4\pi^2}M^2,
\end{equation}
for large $M$.
This follows from the expression for
the overlap of two coherent states with complex amplitudes 
$\alpha$ and $\beta$
\begin{equation}
\langle\alpha|\beta\rangle=\exp(\alpha^*\beta-|\alpha|^2/2-
|\beta|^2/2).
\end{equation}
Thus, the shorter the interaction time $\tau=1/M$, the more
entanglement the states (\ref{M}) contain, provided the number
of photons is sufficiently large.
\subsection{Decoherence due to photon absorption}
The decoherence of a mode due to photon absorption losses
can be described by a parameter $\eta$
that gives the fraction of photons surviving the absorption process.
The loss process is determined by its effect on coherent states by
\begin{equation}
|\alpha\rangle\otimes|0\rangle_E\rightarrow
|\sqrt{\eta}\alpha\rangle\otimes|\sqrt{1-\eta}\alpha\rangle_E,
\end{equation}
where $E$ refers to the environment, which will be traced out.
We have a bipartite state on two modes,
and when we assume both modes to decohere
in the same way, the pure state
$|\Phi\rangle$ is turned into a mixture of the form
\begin{eqnarray}\label{rho}
\rho=\sum_{q=1}^{M}\sum_{p=1}^{M}
f_qf_p^*
(\langle \Psi_p|
\Psi_q\rangle)^2
\nonumber\\
|\tilde{\Phi}_q\rangle
\langle \tilde{\Phi}_p|\otimes
|\tilde{\Phi}_q\rangle
\langle \tilde{\Phi}_p|.
\end{eqnarray}
Here we defined two new states, related to those defined
in  (\ref{Phi}). First, due to decoherence
the amplitude $\alpha$ is reduced by a factor of $\eta$,
and hence we used the states
\begin{equation}\label{Phit}
|\tilde{\Phi}_q\rangle=|\sqrt{\eta}\alpha\exp(-2\pi iq/M)\rangle,
\end{equation}
to expand $\rho$ in. Second,
the state of the environment can be expanded, likewise,
in terms of the states
\begin{equation}\label{PhiE}
|\Psi_q\rangle=|\sqrt{1-\eta}\alpha\exp(-2\pi iq/M)\rangle.
\end{equation}
After tracing out the environment the overlaps of such states
appear in the expression for $\rho$, hence
the appearance of the factor $(\langle \Psi_p|
\Psi_q\rangle)^2$ in (\ref{rho}).
\subsection{Decoherence of entanglement}
In order to determine the entanglement in the state $\rho$
as a function of $\eta,\alpha,M$, we use, as announced, 
the logarithmic negativity
as measure of entanglement \cite{vidal}.
We first have to express the density matrix in some orthogonal
basis. This can be accomplished as follows: first define the $M$-by-$M$
Gram matrix $G$ by its entries
\begin{equation}
G_{pq}=\langle \tilde{\Phi}_p|\tilde{\Phi}_q\rangle.
\end{equation}
Then suppose we have chosen $M$ orthogonal states $|x_i\rangle$
for $i=1\ldots M$ in terms of which we expand
\begin{equation}
|\tilde{\Phi}_q\rangle=
\sum_{i=1}^M A_{qi}|x_i\rangle.
\end{equation}
Similarly, we  expand the density matrix $\rho$
in terms of the basis formed by the set
\begin{equation}\label{x}
|x_i\rangle\langle x_j|\otimes |x_k\rangle\langle x_l|.
\end{equation}
The matrix $A$ can be expressed in terms of the Gram matrix 
by using the relation
\begin{equation}
G_{pq}=\sum_{j=1}^M A_{qj}A_{jp}.
\end{equation}
Namely, in brief notation, we have
\begin{equation}
A=\sqrt{G^T},
\end{equation}
with the superscript $T$ denoting the transpose. Here we used the fact
that $G$ is
a Hermitian matrix with positive eigenvalues, so that $A$, too, can be
chosen as a Hermitian matrix with positive eigenvalues.
Defining a matrix $C$ by its elements
\begin{equation}
C_{mn,q}=f_qA_{qm}A_{qn}
\end{equation}
we can write the density matrix $\rho$
as
\begin{equation}
\rho=C G_2 C^T,
\end{equation}
where $G_2$ is the matrix defined by
\begin{equation}
(G_2)_{pq}=(\langle \Psi_p|
\Psi_q\rangle)^2.
\end{equation}
The partial transpose of $\rho$
with respect to the first mode
and with respect to the basis (\ref{x})
 is denoted by $\rho^{T_1}$ and is defined through
\begin{equation}
\rho^{T_1}_{kl,mn}=\rho_{ml,kn}.
\end{equation}
The logarithmic negativity of $\rho$, finally,
is given by
\begin{equation}\label{exact}
E_N(\rho)=\log_2 ||\rho^{T_1}||,
\end{equation}
with $||.||$ denoting the trace norm, which for
Hermitian operators is equal to
the sum of the absolute values of
the eigenvalues.
The result (\ref{exact}) is exact and can be calculated
numerically
in a straightforward way by using
standard linear algebra. On the other hand, the size of the matrix
$\rho$ may become too large for large $M$
to be useful in numerical computations. Here we use Matlab,
and whenever $M\leq 50$
we use the exact result (\ref{exact}). In other cases
we can use certain approximations valid 
under the appropriate conditions.
\subsection{Some approximations}
In the case that the states $|\tilde{\Phi}_q\rangle$
are almost orthogonal, we can approximate $E_N$ as follows.
First, the partial transpose of $\rho$ 
is approximated by
\begin{eqnarray}
\rho^{T_1}\approx\sum_{q=1}^{M}\sum_{p=1}^{M}
f_qf_p^*(\langle \Psi_p|
\Psi_q\rangle)^2\nonumber\\
|\tilde{\Phi}_p\rangle
\langle \tilde{\Phi}_q|\otimes
|\tilde{\Phi}_q\rangle
\langle \tilde{\Phi}_p|.
\end{eqnarray}
Then the $M^2$ eigenvalues of this matrix can be written as
\begin{eqnarray}
\lambda_{k,m}^{\pm}&=&\pm 
\frac{1}{M}\exp(-(1-\eta)|\alpha|^2|\exp(2\pi i(k-m)/M)-1|^2)
\nonumber\\
&&{\rm for}\,\,(1\leq k<m\leq M)
\nonumber\\
\lambda_k&=&\frac{1}{M}\,\,{\rm for}\,\, (1\leq k\leq M).
\end{eqnarray}
Hence
\begin{equation}\label{EN1}
E_N(\rho)\approx\log_2[1+F],
\end{equation}
where
\begin{equation}\label{EN2}
F=\sum_{k=1}^{M-1}
\exp(-(1-\eta)|\alpha|^2|\exp(2\pi ik/M)-1|^2).
\end{equation}
We can further approximate this when
\begin{equation}
\epsilon=1-\eta
\end{equation}
is a small number, $\epsilon\ll 1$.
In fact, when not only $\epsilon$ is small but $\epsilon|\alpha|^2$
is small as well, we can simply expand the exponentials
appearing in the expression for $F$ to obtain
\begin{eqnarray}\label{approx1}
E_N(\rho)&\approx& \log_2 [M-
2\epsilon|\alpha|^2(M-1)]\nonumber\\
&\approx& \log_2 M-
\epsilon|\alpha|^2 \frac{2}{\log(2)}\frac{M-1}{M}.
\end{eqnarray}
When $\epsilon$ becomes larger, such that $\epsilon|\alpha|^2$ 
is no longer small,
we can make a different approximation in the case
that relation
(\ref{d}) holds with $\delta\ll 1$ some fixed small number.
In this case, $F$ can be rewritten as
\begin{equation}
F=\sum_{k=1}^{M-1}
\exp(-\epsilon\log(1/\delta)M^2|\exp(2\pi ik/M)-1|^2/(2\pi)^2).
\end{equation}
This sum cannot be performed analytically, but
we can approximate it as
\begin{equation}
F\approx \sum_{k=1}^{M-1}
\delta^{k^2\epsilon},
\end{equation}
which is valid for $\epsilon$ not too small.
Since $\delta\ll 1$ only terms with sufficiently small values of $k$ 
contribute. In particular, if we simply cut off the summation
when $\delta^{k^2\epsilon}$ becomes smaller than some number
$D$, i.e., at $k=\sqrt{\log(D)/\log(\delta)}\sqrt{1/\epsilon}$, 
then the entanglement $E_N(\rho)$
is of the form
\begin{equation}\label{D}
E_N(\rho)\approx \log_2[1+\sqrt{D'/(\epsilon\log(\delta)})].
\end{equation}
We checked this approximation
numerically and 
$D'$ was determined to be around $D'\approx \pi$.
\section{Numerical results}
Here we use the exact result (\ref{exact}) and the
approximate results (\ref{EN1}),
(\ref{approx1}),
and (\ref{D}) to investigate the behavior of entanglement
as functions of various different parameters.

When $\alpha$ is sufficiently large 
and $\eta=1$ one will have $\log_2 M$ ebits of entanglement. 
But how quickly does one lose
that possibly large amount of entanglement when there is a
small amount of decoherence? 
The average number of photons absorbed from each mode
is 
\begin{equation}\label{dn}
\Delta N=\epsilon|\alpha|^2.
\end{equation} 
The change in entanglement from pure to mixed state, 
\begin{equation}
\Delta E_N=E_N(\rho)-E_N(|\Phi\rangle\langle\Phi|),
\end{equation}
is, for {\em very} small losses, proportional to $\Delta N$ and is 
given by $2/\log(2)\approx  2.89$ ebits per photon:
\begin{equation}
\Delta E_N\approx-\frac{2}{\log(2)}\frac{M-1}{M}\Delta N,
\end{equation}
where we used
(\ref{approx1}).
This shows how fragile the initial pure entangled states are
when they contain a large number of photons. Even when a tiny fraction
$\epsilon$ of photons is lost, 
the number of photons lost may still be appreciable if
$|\alpha|^2$ is large, and this may lead to a large
degradation of entanglement.

When $\epsilon$ increases the entanglement loss
per photon decreases. Using (\ref{EN1}) [since we assume $\epsilon|\alpha|^2$
is not small we cannot use (\ref{approx1}); we do assume $\delta\ll 1$ now]
we plot
$d E_N/dN$ as a function of $\log_{10}\epsilon$ in Fig.~1.
\begin{figure}
\includegraphics[scale=0.4]{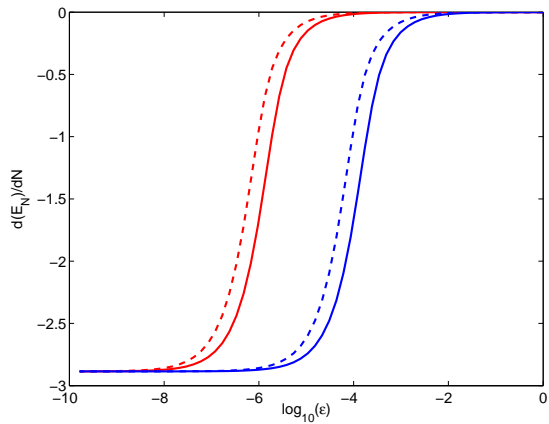}
\caption{Loss of entanglement per photon
$dE_N/dN$ as a function of $\log_{10}\epsilon$.
Two curves on the right correspond to $M=200$, the two 
on the left to $M=2000$.
The solid curves correspond to $\delta=10^{-2}$
and the dashed curves to $\delta=10^{-4}$.
This plot uses the approximation (\ref{EN1}).
}
\end{figure}
The 4 curves plotted in Fig.~1 are shifted versions of one another:
in fact, they are really only functions of $\Delta N=\epsilon|\alpha|^2$.

The approximation (\ref{D}), too,
depends only on $\Delta N$,
not on $\delta$ or $\epsilon$ separately.
In addition the change in entanglement $\Delta E_N$
does not depend
on $M$ either since both $E_N(\rho)$
and $E_N(|\Phi\rangle\langle\Phi|)$
depend on $M$ in the same way. To verify this behavior and to
see whether this behavior extends to parameter regimes
not covered by
the conditions $\delta\ll 1$ and $\epsilon\ll 1$ 
we plot in Fig.~2 the change in entanglement
$\Delta E_N$ as a function
of $\Delta N$ for various different parameters.
For large $M$ we use the approximation (\ref{EN2}),
but for $M=20$ we use the exact result (\ref{exact}).
\begin{figure}
\includegraphics[scale=0.4]{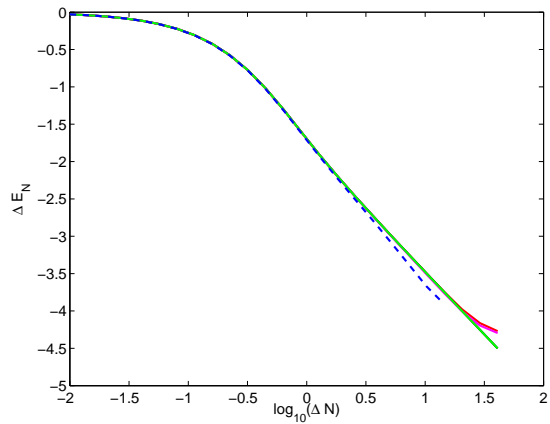}
\caption{Change in
logarithmic negativity, $\Delta E_N$
as a function of the change in number of photons,
$\Delta N$. There are in fact 5 curves here,
several of which overlap to make them indistinguishable.
The dashed curve is for $M=20$ and $\delta=0.2$.
The solid, almost overlapping curves,
are for $\delta=0.01$ and $M=20$,
and for $\delta=10^{-4}$ with $M=20$, 200, and 2000.}
\end{figure}
The plot confirms the predicted behavior: for $\delta$ not too large
the change in entanglement is only a function
of $\Delta N$, but not of $M$, or $\epsilon$ or $\delta$ independently.
For $\delta=0.2$ (and $M=20$)
there is only a small deviation from that universal behavior.

Fig.~3 plots, in contrast, not the change in entanglement but
the entanglement $E_N(\rho)$
left after decoherence. The parameter regime considered
is very different as 
$\epsilon$ is no longer assumed very small.
We do assume $\delta\ll 1$ so that we can use the approximation
(\ref{EN1}). From the approximation (\ref{D})
to that approximation
we see that $E_N(\rho)$ no longer depends on $M$, provided
$M$ is not too small.
That behavior, too, is confirmed in Fig.~3, where curves for $M=20$
(recall that whenever $M\leq 50$ the exact result (\ref{exact}) is used)
are seen to overlap with one for $M=20,000$.
\begin{figure}
\includegraphics[scale=0.4]{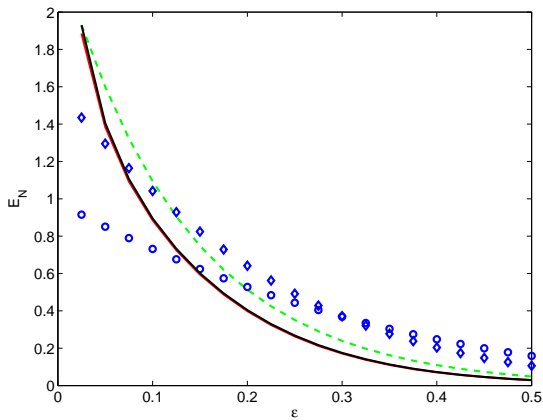}
\caption{
Logarithmitic negativity $E_N(\rho)$
as a function of
$\epsilon=1-\eta$. There are two solid curves almost completely overlapping
that correspond to $M=20$ and $M=20000$, demonstrating
that the amount of entanglement left after decoherence is insensitive to
$M$ when $M$ is not too small. The dashed curve corresponds to
$M=5$, the circles to $M=2$, the diamonds to $M=3$.
}
\end{figure}
For smaller values of $M$ there is a clear dependence on $M$,
as is illustrated for $M=2,3,5$.
From Fig.~3 one notices that at reasonable (i.e. realistic)
values of $\epsilon$
the amount of entanglement left is less than one unit,
even if we started
out with a state with $\log_2 M\approx 14.3$ for $M=20,000$.
However, that result is not optimized for the number of photons
$|\alpha|^2$ in the initial state.
In particular, since we chose $\delta$ to be small,
$|\alpha|^2$ is roughly equal to $M^2$. 

In order to see whether smaller numbers 
of photons may leave more entanglement after decoherence (since one expects
the decoherence to have a less strong influence when there are fewer photons
even if the initial pure state does not
quite have the full $\log_2 M$ ebits of entanglement),
we plot in Fig.~4 $E_N(\rho)$ as a function
of $|\alpha|^2$ for various values of $M$.
\begin{figure}
\includegraphics[scale=0.4]{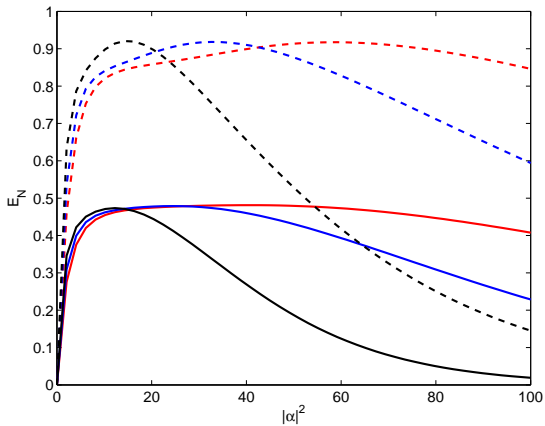}
\caption{
Logarithmitic negativity $E_N(\rho)$
as a function of
$|\alpha|^2$. 
There are three pairs of curves, each pair consisting of
a solid curve, correpsonding to
$\eta=0.7$, and a dashed curve (for $\eta=0.49$). 
The three pairs correspond to $M=20$, 30, and 40, with the maxima
of the curves with larger $M$ moving to larger $|\alpha|^2$.
}
\end{figure}
Here we cannot use the approximation (\ref{EN1}) and hence
use the exact calculation, which, as mentioned before, 
does restrict the calculation to
$M\leq 50$. The plot shows that the optimum
amount of entanglement remaining after
decoherence is almost independent of the value of $M$, thus confirming the
findings from Fig.~3. 
For example, for $\eta=0.7$ one is left with 
about 0.92 ebits of entanglement
for the optimum values of $|\alpha|^2$
almost irrespective of how much entanglement the initial pure state had.
On the other hand, there is a strong dependence
on $|\alpha|^2$.
Finally,
the plot also shows that the decohered states become more robust
against photon absorption losses. For, even if one started
out with a very large amount of (pure-state) entanglement,
losing 30\% of the photons ($\eta=0.7$) leaves one with at best around 0.9
ebits of entanglement. But a further loss of 30\% (so that $\eta=0.49$)
of the photons merely decreases the entanglement by a factor of 2.

In Fig.~5 we plot the same function but for smaller values of
$M$. As it turns out, if one fixes $\eta=0.7$
the largest amount of entanglement one can be left with,
starting out with multi-dimensional entangled coherent states,
is in fact $E_{{\rm max}}\approx 0.95$ achieved by taking
$M=8$ and $|\alpha|^2\approx 2.86$.
\begin{figure}
\includegraphics[scale=0.4]{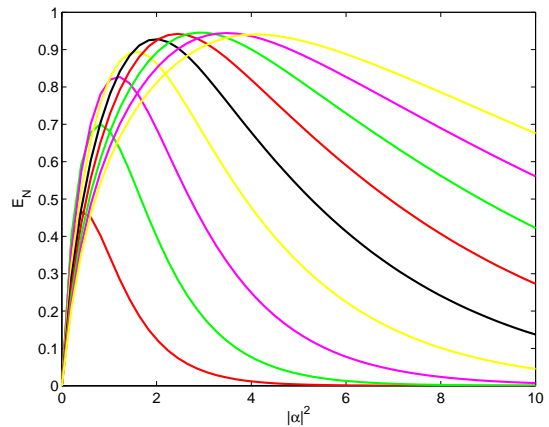}
\caption{Similar to Fig.~4:
$E_N(\rho)$ as a function of
$|\alpha|^2$
for $\eta=0.7$ and $M=2,3,4,5,6,7,8,\ldots 10$,
where the maxima of the curves move to larger
values of $|\alpha|^2$ for larger $M$.
}
\end{figure}

\section{Conclusions}
We considered the decoherence of multi-dimensional entangled
coherent states due to photon absorption losses.
Those states are bipartite entangled states of two modes,
and are described by two independent parameters.
One is $\alpha$, such that $|\alpha|^2$ gives
the average number of photons in each mode.
The other
is $M$, an integer indicating the maximum possible
amount of
entanglement between the two modes, $\log_2 M$.
The amount of photon absorption is governed by an additional
 parameter, $\epsilon$,
the fraction of photons absorbed from each mode.
We calculated the amount of entanglement left after decoherence
as a function of $\alpha$, $\epsilon$, and $M$.
One useful quantity that can be constructed from the above parameters is
$\Delta N=\epsilon|\alpha|^2$, the number of photons
absorbed from each mode. Another useful quantity
is $\delta$ defined as $\delta=\exp(-|\alpha|^2 |\exp(2\pi i/M)-1|^2)$,
which determines the overlap between
two coherent states of amplitude $\alpha$ and phase difference
$2\pi/M$.
In terms of those quantities the conclusions are 
\begin{enumerate}
\item For large $M$ the entanglement disappears at a rate
given by $2/\log(2)\approx 2.89$ 
ebits per photon as long as $\Delta N\ll 1$.
\item For large $M$, small $\delta$ and
not too large values of $\Delta N$
the change in entanglement is a function
of $\Delta N$ alone.
\item For large $M$ and $\Delta N$, and for small $\delta$ 
it is the remaining amount of entanglement
that is a function of $\Delta N$ alone.
\item In general there is an optimum number of photons
$|\alpha|^2$ for fixed $\epsilon$ and $M$
that maximizes the amount of entanglement left after decoherence.
For that optimum, $\delta$ is not very small (around 0.2).
\end{enumerate}
As an example of point 4, we found that
almost one ebit of entanglement can be obtained with 30\%
photon absorption by using a little less than 3 photons
per mode for the initial pure state
and $M=8$. The corresponding value of $\delta$ is
about 0.19. That is the maximum amount possible,
but almost the same amount of entanglement (at least 95\% of the optimum)
can be obtained by using almost any value of $M$ and the
appropriate value for $|\alpha|^2\approx M^2/30$.

\end{document}